\documentclass[conference]{IEEEtran}
\usepackage{fancybox}
\usepackage[ruled,vlined]{algorithm2e}
\usepackage{graphicx}
\usepackage{paralist}
\usepackage{tabularx}
\usepackage[hyphens]{url}
\usepackage{balance}
\usepackage{multirow}
\usepackage{multicol}
\usepackage{pbox}
\usepackage{mathtools}
\usepackage{cite}
\usepackage{adjustbox}
\usepackage[TABBOTCAP]{subfigure}
\usepackage{url}
\usepackage{balance}
\usepackage{amsmath}
\usepackage{verbatim}
\usepackage{float}
\usepackage{colortbl}
\usepackage{color}
\usepackage{amssymb}
\usepackage{ifthen}
\usepackage{pifont}
\usepackage{tikz}
\newcommand*\circled[1]{\tikz[baseline=(char.base)]{
            \node[shape=circle,draw,inner sep=0.5pt] (char) {#1};}}

\newcommand{\CrashScopesp}{{\sc CrashScope~}}
\newcommand{\CrashScope}{{\sc CrashScope}}
\newcommand{\CrashScopes}{{\sc CrashScope's~}}

\include{macro}

\begin{document}

\title{CrashScope: A Practical Tool for Automated Testing of Android Applications}

\author{\IEEEauthorblockN{Kevin Moran$^1$, Mario Linares-V\'asquez$^2$, Carlos Bernal-C\'ardenas$^1$, Christopher Vendome$^1$, and Denys Poshyvanyk$^1$}
\IEEEauthorblockA{
$^1$College of William \& Mary, Williamsburg, VA, USA\\
$^2$Universidad de los Andes, Bogot\'a, Colombia\\
kpmoran@cs.wm.edu, m.linaresv@uniandes.edu.co, \{cebernal, cvendome, denys\}@cs.wm.edu}}

\maketitle
\vspace{-0.5cm}
\begin{abstract}
Unique challenges arise when testing mobile applications due to their prevailing event-driven nature and complex contextual features (e.g. sensors, notifications).  Current automated input generation approaches for Android apps are typically not practical for developers to use due to required instrumentation or platform dependence and generally do not effectively exercise contextual features. To better support developers in mobile testing tasks, in this demo we present a novel, automated tool called \CrashScope.  This tool explores a given Android app using systematic input generation, according to several strategies informed by static and dynamic analyses, with the intrinsic goal of triggering crashes. When a crash is detected, \CrashScopesp generates an augmented crash report containing screenshots, detailed crash reproduction steps, the captured exception stack trace, and a fully replayable script that automatically reproduces the crash on a target device(s).  Results of preliminary studies show that \CrashScopesp is able to uncover about as many crashes as other state of the art tools, while providing detailed useful crash reports and test scripts to developers. Website: www.android-dev-tools.com/crashscope-home Video url: https://youtu.be/ii6S1JF6xDw\end{abstract}


\vspace{-0.2cm}
\section{Introduction}
\label{sec:intro}
\vspace{-0.15cm}
Proliferation in the mobile hardware and application marketplace is increasingly driven by a user base who prefers to carry out computing tasks in the convenient setting of a smartphone or tablet.   However, the gesture-driven nature of mobile apps has given rise to new challenges encountered by programmers during development and maintenance, specifically with regard to testing and debugging \cite{23Joorabchi:ESE13}.   One of the most difficult \cite{3Bettenburg:FSE08} and important maintenance tasks is the creation and resolution of bug reports\cite{36Gu:ICSE10}.  
Reports concerning application crashes are of particular importance to developers, because crashes represent a severe fault that is directly user facing and immediately impacts an app's usability.   If an app is not behaving as expected due to crashes, missing features, or other bugs, nearly half of users will likely to abandon the app for a competitor \cite{app-abandonment} in a marketplace such as Google Play. Therefore, to ensure their app's success, developers need automated support to uncover, reproduce, and fix crashes.

While significant progress has been made in the area of testing and automatically generating inputs for mobile applications, the available tools generally exhibit some noteworthy limitations: (i) previous approaches lack the ability to provide detailed, easy-to-understand testing results for faults discovered during automatic input generation, leaving the developer to sort through and comprehend stack traces, log files, and non-expressive event sequences \cite{Choudhary:ASE15}; (ii) most approaches for automated input generation are not practical for developers, typically due to instrumentation or difficult setup procedures.  This is affirmed by the fact developers typically prefer manual over automated testing approaches \cite{Kochhar:ICST15}; (iii) no approach combines different GUI-exploration strategies, targeted testing of contextual features, and multiple policies for user text input in a single holistic approach.  These shortcomings contribute to the low adoption rate of automated testing approaches and the preference of manual and scripting based testing techniques.

Motivated by these current issues developers face in adopting automated testing tools in their workflow, we designed and implemented \CrashScope, a practical system that automatically discovers, reports, and reproduces crashes for Android applications.  \CrashScopesp explores a given app according to a set of several input generation strategies and produces expressive crash reports with explicit steps for reproduction in an easily readable natural language format.  This approach requires only an \texttt{.apk} file (and optionally source code) and an Android emulator or device to operate and \textit{does not} rely on instrumentation of the subject apps or the Android OS.  \textit{The entirety of the \CrashScopesp workflow is automated, requiring no developer intervention, other than reading produced reports}. The differing exploration strategies supported by our tool are aimed at eliciting crashes from Android apps and include automatic text generation capabilities based on the context of allowable characters for text entry fields, and targeted testing of contextual features, such as the orientation of the device, network interfaces, and sensors.  We specifically tailored these features to test the common causes of app crashes as identified by previous studies~\cite{Zaeem:ICST2014,Liang:ICMCN2014}. During execution, \CrashScopesp captures detailed information about the subject app, such as the inputs, screenshots, GUI information, exceptions, and crashes.  This information is then translated into detailed crash reports and replayable scripts, drastically reducing the burden on developers attempting to fix faults.

\vspace{-0.15cm}
\section{The CrashScope Automated Testing Tool}
\label{sec:approach}
\vspace{-0.15cm}
\CrashScopesp addresses the general limitations of existing tools in several ways that illustrate the novelty of the approach: (i) \CrashScopesp is able to automatically generate expressive bug reports (and replayable scripts) for exceptions or crashes uncovered; (ii) the approach is practical requiring only an \texttt{.apk} file to operate; (iii) the exploration strategies enable targeted testing of contextual features and adaptive text input generation; and (iv) the tool is \textit{app-crash-resilient}; it can detect a crash and continue testing the unvisited components and states of the GUI after handling the crash.  \CrashScopes target user base is mobile application developers and testers who wish to incorporate a practical automated testing tool into their workflow.  To make the tool easy to use and to allow for efficient integration into a typical developer's workflow we have developed a could-based Java Web application which encapsulates the \CrashScopesp tool. This web application includes an \textit{Execution Engine} back-end, which performs the static and dynamic analyses required for testing, and a user-friendly front-end making it easy to submit apps for testing.  This architecture ensures the flexibility, scalability, and extensibility of \CrashScopesp as a web-based tool.

\vspace{-0.15cm}
\subsection{CrashScope Design}
\label{subsec:approach-overview}

\begin{figure}[t]
\centering
\vspace{-0.7cm}
\includegraphics[width=\linewidth]{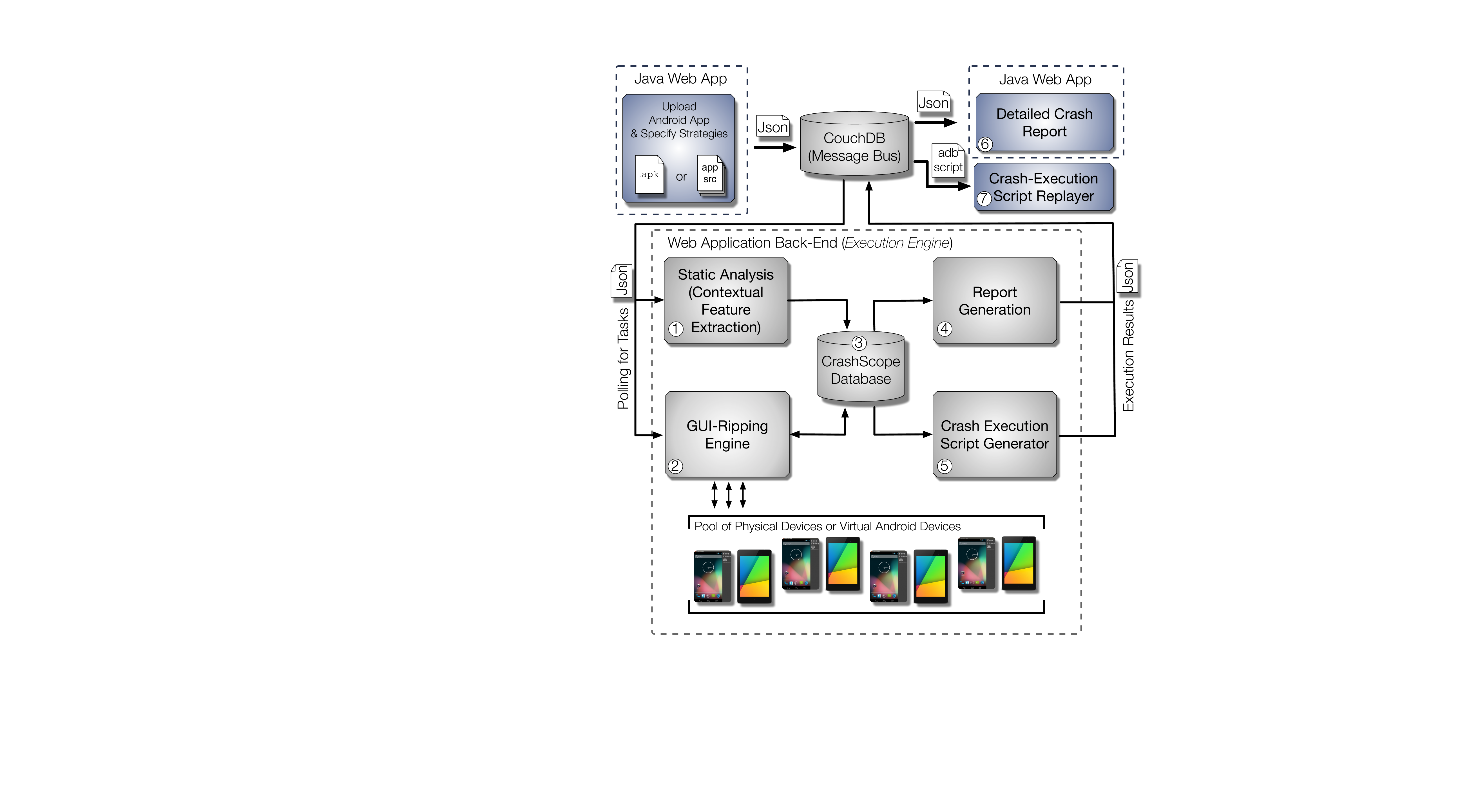}
\vspace{-0.7cm}
\caption{\CrashScopesp Design}
\label{Design}
\end{figure}

\begin{figure}[t]
\vspace{-0.7cm}
\centering
\includegraphics[width=\columnwidth]{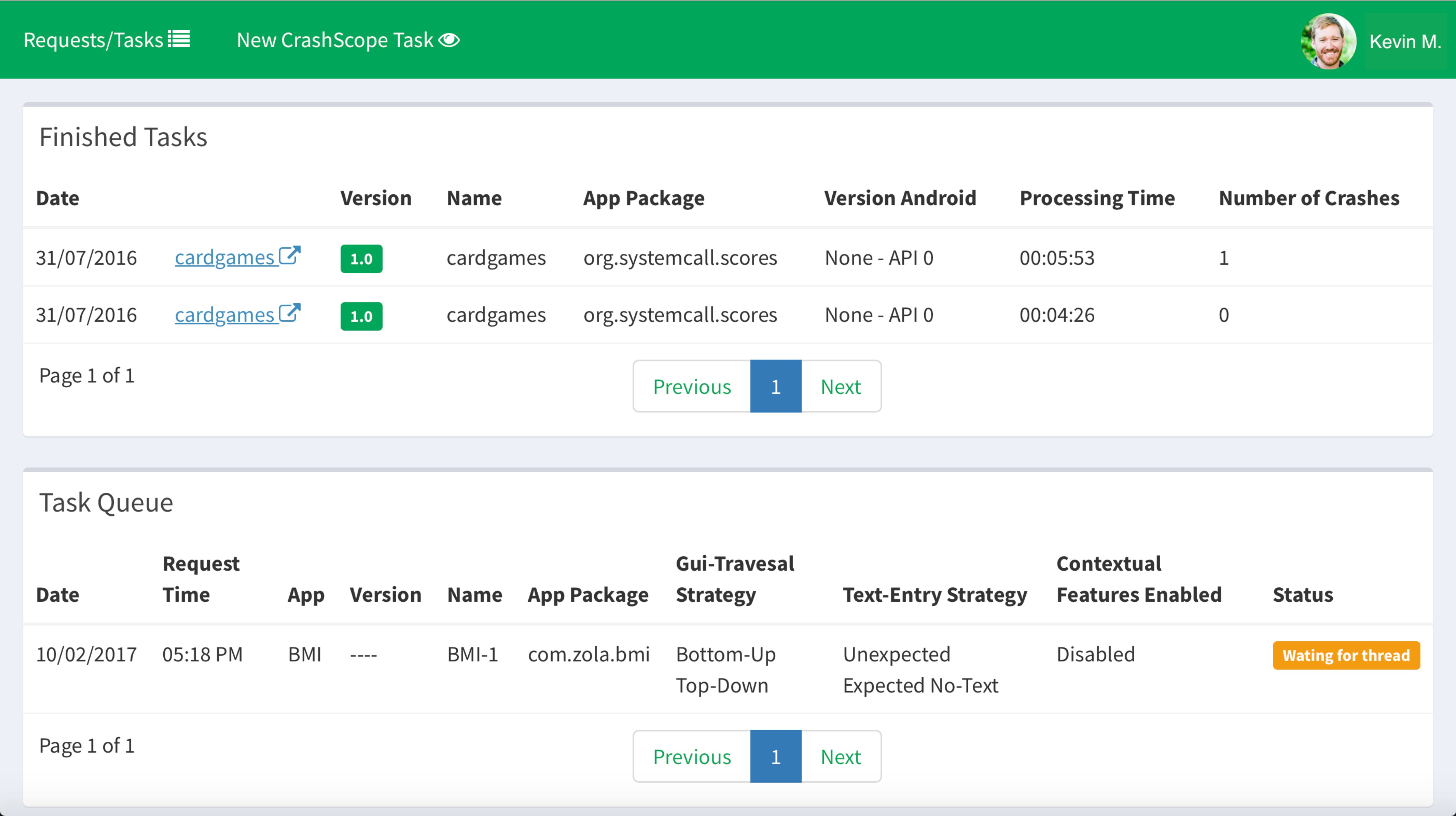}
\vspace{-0.7cm}
\caption{\CrashScopesp Web App Experiments Dashboard}
\label{fig:web}
\end{figure}

	The overall architecture of \CrashScopesp is illustrated in Figure \ref{Design}. The first step in the workflow of the tool is to submit an application for testing by using the \textit{New Task} form of the web application to upload an app \texttt{.apk} file (and optionally src code) and choose the configuration of strategies that \CrashScopesp will use to explore the app. Once the task has been initiated by the developer the progress can then be viewed in the \textit{Testing-Dashboard} (Figure \ref{fig:web}).  On the \textit{Testing-Dashboard} page a developer can view submitted tasks and execution progress of those currently running (e.g. bottom portion of Fig. \ref{fig:web}). Additionally,  statistics, including the running time, app information and \# of crashes discovered, for completed tasks can be viewed (e.g. top portion of Fig. \ref{fig:web}). After a task is initiated, the web app sends the task information, including the \texttt{.apk} file, to an instantiation of  \texttt{CouchDB} (acting as a message queue) through a \texttt{json} document. The backend \textit{Execution Engine} periodically polls for new tasks on the queue, initiating the testing process when a message is received.   The first step the \textit{Execution Engine} performs in the testing process is static analysis of the submitted app to identify contextual features (Figure \ref{Design}-\circled{1}).  To do this it either utilizes the user-uploaded source code of the app, if it was provided, or attempts to decompile the app using a combination of tools\footnote{ \texttt{apktool}, \texttt{dex2jar} and \texttt{jd-cmd}}.  It then detects Activities (e.g. screens) that are related to contextual features in order to target the testing of such features. In other words, \CrashScopesp will only test contextual app features (e.g., network on/off) if it identifies feature instances in app code (Sec. \ref{subsec:context-features}).

Next, the \textit{GUI Ripping Engine} (Figure \ref{Design}-\circled{2}) systematically executes the app using the user-specified combination of strategies (Section \ref{subsec:strategies}), including enabling and disabling the contextual features (if run on an emulator) at the pertinent screens identified by static analysis.  If during the execution, uncaught exceptions are thrown, or the app crashes,  dynamic execution information is saved to \CrashScope's database (Figure \ref{Design}-\circled{3}), including detailed information regarding each event (e.g. touch-event, text entry, sensor value change) performed during the systematic exploration.  This execution can be carried out on a large number of concurrently running physical devices or Android Virtual Devices (AVDs) which are instantiated using \texttt{android-x86} images and VirtualBox.
	
	After the execution data has been saved to the \CrashScopesp database, the \textit{Natural Language Report Generator} (Figure \ref{Design}-\circled{4}, Section \ref{subsec:reports}) parses the database and processes the information for each event of all execution sequences that ended in a crash, and sends this information via \texttt{json} back to \texttt{CouchDB}.  Then the Java web app generates an HTML based natural language crash report with expressive reproduction steps (Figure \ref{Design}-\circled{6}) for the user.  In addition, the \textit{Crash Script Generator} (Figure \ref{Design}-\circled{5}, Section \ref{subsec:scripts}) parses the database and extracts the relevant information for each step in a crashing execution in order to create a replayable script containing \texttt{adb input} commands and markers for contextual state changes.  The \textit{Script Replayer} (Figure \ref{Design}-\circled{7}, Section \ref{subsec:scripts}) is a small Java program that is able to replay these scripts.
		
\vspace{-0.15cm}
\subsection{Extracting Activity and App-Level Contextual Features}
\label{subsec:context-features}

	\CrashScopesp uses Abstract Syntax Tree (AST) based analysis to extract the API-call chains that are involved in invocations of contextual features. In particular, it detects Android API calls related to network connectivity and sensors (i.e., Accelerometer, Magnetometer, Temperature Sensor, and GPS). Because the API calls might not be executed directly by the code associated with a class implementing an Activity (which comprises a screen in Android), \CrashScopesp performs a call-graph analysis to extract paths ending in a method invoking a contextual API.  Because certain API calls may not be traceable through a back-propagated call-chain (e.g., sensors or network connectivity implemented as a service), \CrashScopesp employs two granularities for testing contextual features: activity (screen-) level and app-level. If the API analysis idicates that an app implements a certain contextual feature, but this feature is not able to be linked back to a particular screen by call-chain analysis, then the feature is tested at the ``app-level" or at each screen of the app.  Rotatable screens are located through analysis of an app's \texttt{AndroidManifest.xml} file. 
	
\vspace{-0.15cm}
\subsection{Exploration of Apps \& Crash Detection}

	To explore an app, \CrashScopesp dynamically extracts the GUI hierarchy of each app screen as it is visited during exploration (using \texttt{uiautomator}) and identifies the clickable and long-clickable components to execute, as well as available components for text inputs (e.g., EditText boxes). The (long-) clickable components and are added to a dynamic \textit{component-stack} as they are observed on each screen, and executed according to the order of the stack.  Currently, the \textit{Ripping Engine} supports the \textit{tap}, \textit{long-tap}, and \textit{type} events. If text entry fields are available in a particular app screen, then each text box is filled in before each (long-) clickable component on the screen is exercised.  Additionally, each transition from one unique activity to another is recorded in a \textit{transition graph} along with the events that led to transition.  The component and transition graph allow \CrashScopesp to build an accurate real time model of the app during execution.  
		
	Text entry from the user is a major part of functionality in many Android apps, therefore, \CrashScope's \textit{GUI Ripping Engine} employs a unique text input generation mechanism.  \CrashScopesp detects the type of text expected (e.g., phone \#, email address) by a text field, by querying the keyboard type associated with the text field \cite{android-keyboard} using the  \texttt{adb shell dumpsys input\_method} command. Once the type of expected input is detected, \CrashScopesp employs two hueristics to generate text inputs: \textit{expected} and \textit{unexpected}.	
 The \textit{expected} hueristic generates a string within keyboard parameters without any punctuation or special characters, whereas the \textit{unexpected} heuristic generates random strings with all of the allowable special characters for a given keyboard type.
	
   In addition to the text input generation strategies, \CrashScopesp can generate touch inputs according one of several GUI-exploration strategies.  Currently, the tool traverses the GUI hierarchy in a depth-first-like manner either from the bottom of the hierarchy up or from the top of the hierarchy down (controlled by the order components are placed on the \textit{component stack}).  The rationale for having two such strategies is to generally mimic what a user would do (i.e., executing GUI events without a predefined order).  However, \CrashScopesp is extensible to other types of exploration approaches including random-based or biased-random approaches. To detect and capture exceptions, \CrashScopesp filters the logcat for uncaught exceptions related only to the app being tested.  To detect crashes, \CrashScopesp checks for the appearance of the standard Android crash dialog.  If a crash is encountered, the execution information is logged to the database. However, because the \textit{GUI-Ripping Engine} maintains a \textit{transition graph} and stack of unvisited components, the execution can continue towards remaining program paths.

\vspace{-0.2cm}
\subsection{Testing Apps in Different Contextual States}	

	When GUI-Ripping begins, \CrashScopesp first checks for app-level contextual features that should be tested according to the exploration strategy.  As each activity is visited, the \textit{GUI-Ripping Engine} checks if contextual features should be enabled/disabled and sets feature values according to the current strategy.  The testing of contextual features works \textit{only on} emulators or AVDs using telnet commands.  While the telnet commands do support turning on/off the network for an emulator, they do not support the enabling/disabling of sensors (Accelerometer, Magnetometer, GPS, Temperature Sensor), but it is possible to mock values for these sensors.  Therefore, to test for sensor related features in adverse conditions, the network connection is disabled, and unexpected (e.g. highly infeasible) values are set for the other sensors such as the GPS and Gyroscope.
		
\vspace{-0.2cm}
\subsection{Multiple Execution Strategies}
\label{subsec:strategies}

	One of \CrashScopes most powerful features is its ability to explore an app according to several different strategies through combinations of its various supported testing features.  These strategies stem from three major feature heuristics: 1) the GUI-traversal strategy (e.g. dfs-top-down, dfs-bottom-up, or random-based), 2) the method by which inputs are generated for user text entry fields (\textit{no text}, \textit{expected text}, \textit{unexpected text}), and finally, 3) enabling or disabling the testing of adverse contextual states (e.g., if an activity is found to have utilized the network, should it be turned on or off?).  Different combinations of these strategies have the potential to uncover different types of app crashes.  By running an app through all combinations of these three feature heuristics in different strategies, \CrashScopesp can effectively test for different types of commonly inducible crashes.  These strategies can also be parallelized by running several combinations for an app concurrently on a group or a cloud of emulator instances (e.g. using the \textit{Execution Engine}) further reducing testing overhead.

\vspace{-0.15cm}
\subsection{Generating Expressive, Natural Language Crash Reports}
\label{subsec:reports}

\begin{figure}
\vspace{-0.7cm}
\centering
\includegraphics[width=\columnwidth]{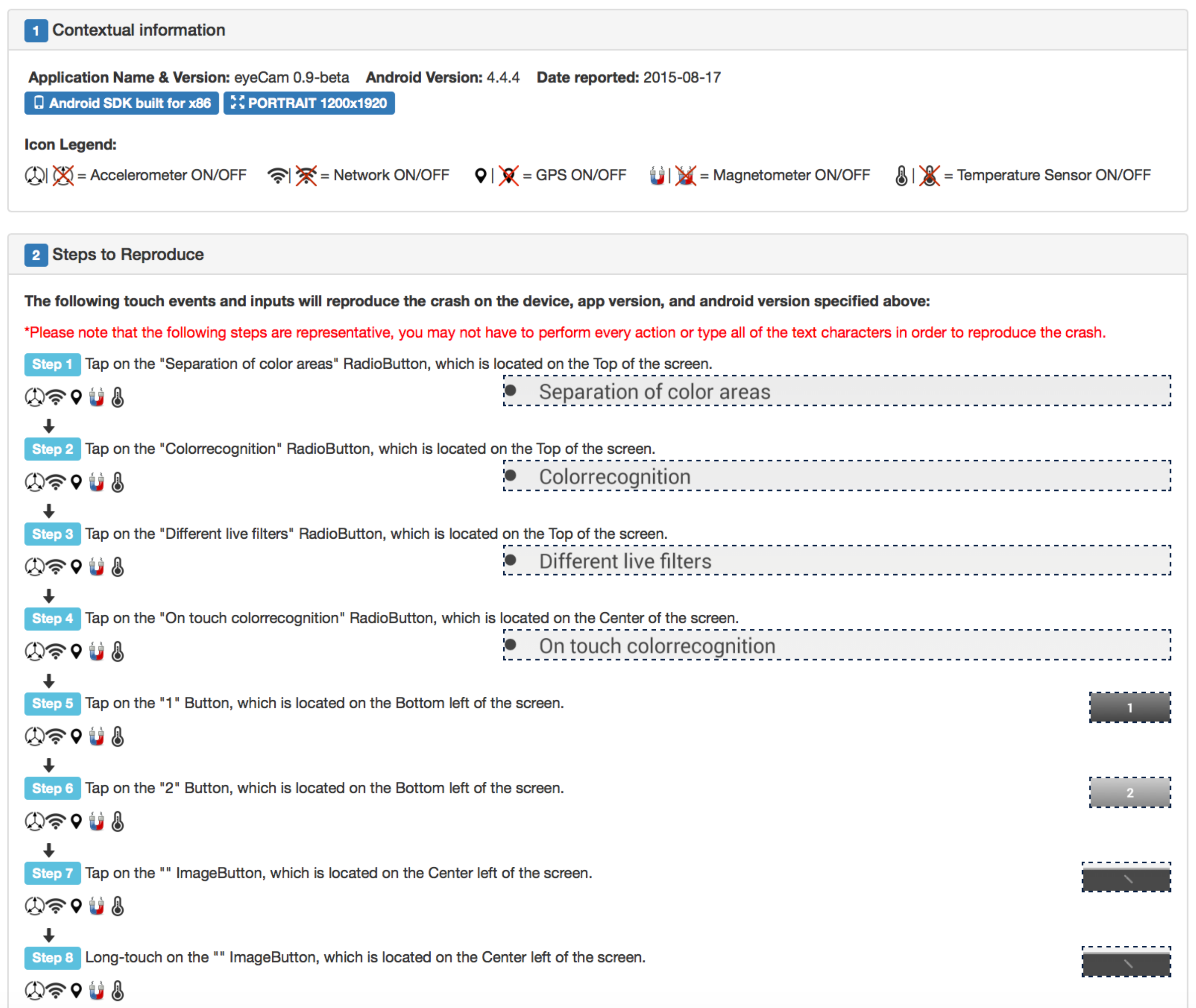}
\vspace{-0.7cm}
\caption{Example \CrashScopesp Report}
\vspace{-0.1cm}
\label{steps}
\end{figure}

	\CrashScopesp generates a Crash Report (Figure \ref{Design}-\circled{6}) that contains four major types of information: 1) general information including the app name and version, the version of the Android OS, a legend of icons that indicates the current contextual state of the app in the reproduction steps, the device, and the screen orientation and resolution when the crash occurred; 2) natural language sentences that describe the steps to reproduce a crash using detailed information about the GUI events and contextual states for each step (Figure \ref{steps}); 3) an app's screen flow that highlights the component interacted with on each screen in the execution scenario for a particular crash; (4) a pruned stack trace containing only the app exceptions that occurred during execution.  Once the execution engine reports that all strategies for a submitted task are completed, it sends a message to  \texttt{CouchDB} via \texttt{json}.  The message is retrieved by the Java Web App  and the results and crash reports for the current task are made viewable in the \textit{crash-report} page.

\vspace{-0.15cm}
\subsection{Generating \& Replaying Reproduction Scripts}
\label{subsec:scripts}
\vspace{-0.1cm}
The  \textit{Crash Script Generator} (Figure \ref{Design}-\circled{5}), parses the saved execution information from the \CrashScopesp database and generates replayable scripts containing \texttt{adb input} commands for touch and text inputs and markers for changes in contextual states.  The scripts are generated by parsing the database for all of the GUI events associated with each step in a particular execution.  Then, the coordinates of each component that were recorded during the systematic exploration of the app are parsed and the center coordinates are extrapolated based on each component's size.  These coordinates are used to generate \texttt{adb input} commands to reproduce the GUI event.  This approach relies on our previous work in replaying events of test sequences in Android apps \cite{Linares:MSR15}. 
The scripts can be replayed by the  \textit{Script Replayer} (Fig. \ref{Design}-\circled{7}), a small java program that takes the recorded script as input, and replays the \texttt{adb} commands and state change markers (e.g. ``WiFi\_off").

\vspace{-0.3cm}
\section{Evaluation}
\vspace{-0.15cm}
\label{sec:evaluation}

To evaluate CrashScope (See full details in \cite{Moran:ICST16}), we performed two empirical studies aimed at investigating both the crash-finding ability of the tool, as well as the reproducibility and readability of the generated reports.  First, in the \textit{Crash Discovery Study}, we compared \CrashScopesp against 4 state-of-the-art automated input generation tools for Android (and the standard Monkey tool) on 61 open source subject applications by utilizing a subset of subject apps and tools available in the Androtest testing suite \cite{Choudhary:ASE15}.  Second, in the \textit{Crash Report Study} we evaluated the reproducibility and readability of the natural language reports generated by \CrashScopesp compared to human written reports found in online issue trackers.  In this study we extracted 8 real-world crash reports from online issue trackers linked to apps on the F-Droid open source marketplace, for which \CrashScopesp was able to generate a report.  Then 16 graduate student participants attempted to reproduce a total of 8 bug reports (4 from \CrashScopesp and 4 from the online issue trackers) and subsequently answered survey questions about the two types of anonymized bugs reports. The results of these studies indicate that (i) \textit{\CrashScopesp is about as effective at uncovering crashes as other state of the art automated input generation tools for Android while reducing the number of false positives due to instrumentation required of other tools} and (ii) \textit{reports generated by \CrashScopesp are as reproducible as human written reports and are more readable and useful from a developer's perspective.}

\vspace{-0.25cm}
\section{Demo Remarks and Future Work}
\label{sec:concl}
\vspace{-0.15cm}
In this demo, we presented \CrashScope, a novel implementation of a practical automated testing approach for Android applications.  Our tool overcomes several shortcomings of previous approaches by (i) leveraging static analysis to locate and test contextual features in a targeted manner; (ii) exploring an app using several different combinations of strategies; (iii) generating highly useful crash reports and replayable test scenario scripts.  In the future, we aim to investigate techniques to trim bug reports to only necessary steps and improve our systematic exploration strategy by adapting promising emerging approaches in model-based GUI testing and static analysis.  Additionally, we aim to distribute \CrashScopesp to the broader Android development community and evaluate its usefulness.
\vspace{-0.1cm}

\balance
\bibliographystyle{abbrv}
\bibliography{ms}

\end{document}